# GAP: Game Theory-Based Approach for Reliability and Power Management in Emerging Fog Computing

Abolfazl Younesi, Mohsen Ansari, Alireza Ejlali, Mohammad Amin Fazli,
Muhammad Shafique, *Senior Member, IEEE*, and Jörg Henkel, *Fellow, IEEE*

*Abstract*— **Fog computing brings about a transformative shift in data management, presenting unprecedented opportunities for enhanced performance and reduced latency. However, one of the key aspects in fog computing revolves around ensuring efficient management of power and reliability. To address this challenge, we have introduced a novel model that propose a non-cooperative game theory-based strategy to strike a balance between power consumption and reliability in decision-making processes. Our proposed model capitalizes on the Cold Primary/Backup strategy (CPB) to guarantee reliability target by re-executing tasks to different nodes when a fault ocurs, while also leveraging Dynamic Voltage and Frequency Scaling (DVFS) to reduce power consumption during task execution and maximize overall efficiency. Non-cooperative game theory plays a pivotal role in our model, as it facilitates the development of strategies and solutions that uphold reliability while reducing power consumption. By treating the trade-off between power and reliability as a non-cooperative game, our proposed method yields significant energy savings, with up to a 35% reduction in energy consumption, 41% decrease in wait time, and 31% shorter completion time compared to state-of-the-art approaches. Our findings underscore the value of game theory in optimizing power and reliability within fog computing environments, demonstrating its potential for driving substantial improvements.**

## I. INTRODUCTION

THE exponential growth of the Internet of Things (IoT) devices will lead to an overwhelming daily data generation of approximately 2.5 quintillion bytes in 2022 [1]. By May 2022, the number of connected IoT devices had already reached nearly 14.4 billion, projected to reach 25 billion by 2025 [1]. Consequently, the substantial volume of generated data requires storage and analysis capabilities. Cloud computing is utilized for this purpose, enabling the execution of various applications with diverse objectives [2][3]. However, cloud computing faces challenges such as network congestion due to high data transmission delays and limited capacity [4]. Additionally, there is a growing demand for real-time computing, which is essential for numerous time-sensitive IoT applications today [5][6][32].

Fog computing was introduced in 2012 and advocated by the OpenFog consortium [8][9]. Fog computing establishes a distributed computing environment near the end user, allowing data collection and processing close to IoT devices, ensuring low latency, high mobility, and real-time capabilities [6][7][31][39][41]. However, fog computing systems face inherent conflicts between optimizing energy management and maintaining high reliability, crucial for sustaining device operation and ensuring dependable performance.

Examples of fog computing applications include distributed wireless networks, self-driving vehicles, and healthcare [7] [11][28][29][30]. Furthermore, fog computing encounters challenges in resource and power management, task scheduling (See Fig. 1), federation, mobility, and fault tolerance [6][7][11][31]. Power management holds tremendous significance for IoT devices, and developing low-power designs has long been a priority for IoT developers [13]. Its objective is to minimize devices' dynamic and static power consumption through various techniques and methodologies, thereby optimizing battery life [13][27]. Moreover, fog computing systems can be designed to operate efficiently and effectively in uncertain environments by considering realistic system components and their interactions [6][11][16].

### A. State-of-the-Art and Their Limitation

With its limited resources, fog computing necessitates efficient task scheduling to optimize resource utilization. Existing task scheduling approaches in fog computing environments have been presented in recent studies [7][16][17][18][19][40]. However, these approaches have focused on specific frameworks, limiting their effectiveness like limited adaptability, scalability, inadequate conflit resolution, and energy overheads [42]. Many of these frameworks do not adequately address the complex trade-offs between multiple competing objectives such as power consumption, reliability, and latency, which are critical in a dynamic fog computing environment. The conflict between energy management and reliability arises because minimizing power consumption often necessitates reducing active components or processing power, which can compromise the system's ability to handle failures and maintain continuous operation. Conversely, ensuring high reliability typically involves deploying redundant systems and robust error-handling mechanisms that increase power usage.

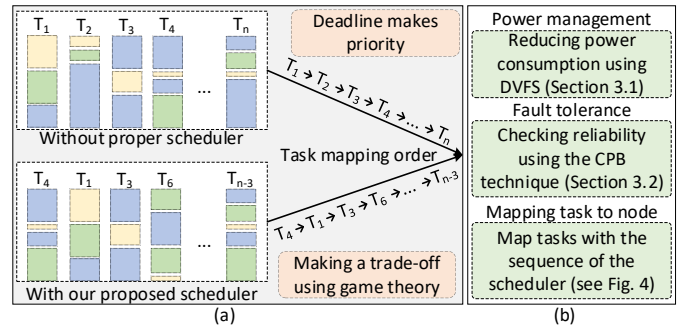

**Fig. 1.** (a) Our proposed task scheduling algorithm achieves a balance between power consumption and reliability using a non-cooperative approach, as opposed to an improper scheduler that does not prioritize deadlines effectively, (b) Achieving optimal or sub-optimal output by following the sequence of modules using non-cooperative game theory principles.

In this paper, we address the crucial parameters of power consumption and reliability by employing game theory to model their trade-offs. Specifically, we focus on decentralized fog computing environments where independent agents interact and make autonomous decisions. Game theory is particularly well-suited for decentralized environments due to its ability to model and analyze the interactions between multiple autonomous agents (or players) with potentially conflicting interests and strategic decision-making. Achieving a delicate balance is essential to avoid overloading tasks, which leads to poor performance and increased power consumption.

Additionally, conflicts may arise during task allocation in a decentralized environment where end users act independently. Hence, game theory is employed in our approach to strike a trade-off between power consumption and reliability while considering resource constraints in task scheduling.

As previously stated, using game theory to solve fog computing challenges enables multiple players to pursue various objectives such as performance, availability, verifiability, reliability, and accountability [10][32]. Within the context of game theory, players aim to understand and predict other players' intentions [12]. Various game models, including zero-sum, non-zero-sum, cooperative and non-cooperative, and dynamic and static games, can be employed depending on the specific challenge and strategies involved. In this paper, we have chosen a non-cooperative game model as it offers a flexible approach to addressing fog computing challenges. This model allows multiple players to act independently and realistically without imposing restrictions on each other [10][12].

Our approach demonstrates effectiveness by addressing the limitations of previous frameworks, specifically in handling the trade-offs between power consumption and reliability. Through evaluations, we show that our game-theoretic model outperforms existing task scheduling methods by better balancing these critical parameters, leading to improved overall system performance in decentralized fog computing environments. This validates the necessity and feasibility of employing game theory in task scheduling within generalized fog computing environments, making our contribution practical.

### B. Our Novel Contributions

Our primary contribution is the development of a novel algorithm that addresses the specific challenge of optimizing power consumption in fog computing environments. This algorithm leverages game theory to navigate the trade-offs between power usage and reliability, especially crucial in resource-constrained fog nodes.

Fog computing faces several challenges, one of the most significant being efficient power management. Due to limited computing resources and processing capabilities, fog nodes must strike a delicate balance between power consumption and ensuring reliable task execution. Our algorithm tackles this issue head-on. To achieve this, we incorporate two key techniques: **Dynamic Frequency-Voltage Scaling (DVFS)**: This allows us to dynamically adjust the voltage and frequency levels of fog nodes, directly impacting power usage [13]. **Cold Primary/Backup (CPB)**: This reliability enhancement strategy ensures that a backup task is ready to execute if the primary task fails, mitigating the risks associated with resource constraints [14]. (Table I details the symbols and notations used in our study).

We adopt a game-theoretic approach due to its effectiveness in analyzing and making strategic decisions in competitive, real-world scenarios [12][32]. By modeling our system as a non-cooperative game, we represent end-user devices as players competing to execute their tasks promptly. The goal is to design an efficient scheduler that optimizes resource allocation to strike the ideal balance between power consumption and reliability while meeting task deadlines [15][18]. Penido's book [15] describes scheduling as a decision-making process commonly employed in various manufacturing and service industries. It involves allocating resources to tasks at specific time intervals to optimize one or more objectives [15]. The objectives revolve around resolving the trade-off between power consumption and reliability while considering timing constraints.

**In a nutshell, the main contributions of our research are as follows** (refer to Fig. 2 for an overview):

- **Game-Theoretic Modeling:** A novel application of game theory to model the power-reliability trade-off in fog computing, leading to optimal or near-optimal solutions.

- **Self-Organized Task Scheduling (GAP):** Introduction of a self-organizing algorithm that efficiently schedules tasks and manages resources in fog environments, considering deadlines.

- **Enhanced Reliability (CPB):** Utilization of the CPB approach to improve system reliability while minimizing power consumption through DVFS.

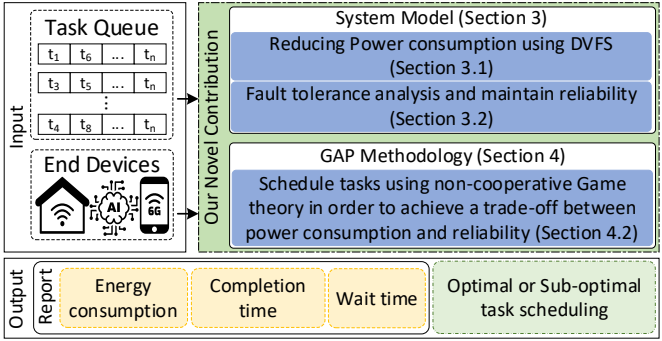

Fig. 2. An overview of our main contributions

- Experimental results demonstrate that our efficient mapping of tasks to resources enables effective resource management, resulting in reduced wait times and lower energy consumption.

### C. Paper Organization

The rest of the paper is organized as follows: The related work and the system model are reviewed in Section 2 and Section 3, respectively. The system, task, power, and fault tolerance models are presented in Section 3. In Section 4, we present our proposed method. In Section 5, the results are presented, and Section 6 concludes the paper.

## II. BACKGROUND AND RELATED WORKS

The existing literature has not adequately addressed the reliability factor, as well as other important considerations such as energy consumption and execution time. When these factors are not properly balanced, conflicts can arise, leading to increased energy or power usage, the need for more processing units, and other inefficiencies. This highlights the critical importance of incorporating reliability and other key factors into the system design process.

To address this gap, this paper proposes a method that jointly optimizes reliability along with energy consumption, execution time, and other relevant factors. The goal is to develop an algorithm that can effectively navigate the trade-offs between reliability and power management, resulting in a more efficient overall system design. A comprehensive comparison of related works is provided in Table II, which highlights the key parameters of each study alongside our proposed method, facilitating a precise evaluation of the different approaches.

To maximize the usage of computing resources in the fog computing environment, [4] proposed a hybrid technique combining GA, PSO, and ACO to optimize resource usage in fog computing, considering factors like construction time, deadline, budget, cost, security, processing resources, load balancing, and energy consumption. The hybrid approach outperformed other heuristic algorithms in terms of energy consumption, execution time, and cost. However, the paper should have also considered reliability when comparing their approach to other methods. This factor could play a vital role in determining its success.

The study [17] proposed a game theory-based approach for scheduling Internet of Things (IoT) services in fog computing environments to address the challenge of delays. The main components of their solution include: 1- Preference functions for IoT devices and fog nodes to rank each other based on criteria such as latency and resource utilization. 2- Centralized and distributed intelligent adaptive algorithms (CIAA) to assist in the scheduling process. The results showed that their proposed method outperformed the Min-Min and Max-Min techniques in terms of execution time, makespan, and resource utilization. The authors evaluated their approach by considering two scenarios: (1) a fixed number of fog nodes and a variable number of IoT devices, and (2) a constant number of IoT devices and a variable number of fog nodes. One of the advantages of this research is the high scalability of the proposed algorithm.

In research [22], a work allocation heuristic technique for software-embedded systems with fog support was proposed. In this paper, they developed a heuristic approach for mixed-integer nonlinear programming (MINLP) to reduce the job completion time factor in the workload distribution between the edge and client sides. The authors presented a three-phase heuristic approach to address the proposed technique's computational complexity issue. By converting MINLP to LP equivalent, the first two phases aim to reduce the computation time and I/O time for each operation. In the third step, storage and computing servers are integrated while considering the transfer time to minimize the transfer time. Finally, the results show that the proposed algorithm has higher efficiency and optimization related to the completion time but needs more scalability.

TABLE I
LIST OF DEFINITIONS

| Symbol | Definition |
|--------|------------|
| $Bw$ | Bandwidth |
| $cb$ | Count of the backup tasks that did not executed in a frequency |
| $c_i$ | Completion time of a task |
| $cp$ | Count of the primary tasks that did not executed in a frequency |
| $CPB$ | Cold Primary/Backup |
| $CT$ | Completion Time |
| $CT_b$ | Completion time of backup task |
| $d_i$ | Deadline of a task |
| $DVFS$ | List of frequencies can be applied on the nodes to run the tasks |
| $ExT$ | Execution time task |
| $ExT_p$ | List of $T_t$ |
| $f_{max}$ | Maximum frequency of resource to execute a task |
| $LCT$ | List of completion time |
| $Lf$ | List of failed tasks |
| $LoT_{dl}$ | List of the tasks Deadline |
| $LR$ | List of Available Resource |
| $Lt$ | List of sorted tasks considering deadline of the tasks |
| $LT_b$ | List of backup tasks |
| $Pp$ | Processing power |
| $Rm$ | Remaining time for backup task |
| $Rt$ | Response time |
| $s_{ij}$ | Start time of the task $i^{th}$ on resource $j$ |
| $ST$ | Task submission time |
| $Tb_i$ | Backup task |
| $t_i$ | Task $i^{th}$ |
| $Tp_i$ | Primary task |
| $T_t$ | Time of Primary task executed until fault |
| $V_{max}$ | Maximum voltage of resource to execute a task |
| $x_{ij}$ | Submitted task $i^{th}$ on resource $j$ |

A four-layer framework supporting workload planning and load balancing (LB) in the fog computing environment was provided in the paper [23]. IoT devices are present in the first layer. Based on the dual fuzzy logic technique, various applications in the second layer are categorized as high priority and low priority. The task size, arrival time, minimum execution time, and maximum completion time are among the inputs taken into account by fuzzy logic methods. Highly prioritized tasks are sent to queue 3, which contains a new fog layout. In this approach, fog nodes are clustered using the K-means clustering algorithm. They evaluate their approach using a real-time program based on schedule, response time, energy consumption, and workload balance ratio.

In the paper [28], the author presents a maximal energy-efficient task scheduling algorithm to develop an energy-efficient fog computing framework for homogeneous fog networks with multiple neighbor helper nodes sharing their computing resources and spectrum access techniques, enabling intelligent IoT applications.

Authers in [30] present a new energy-aware metaheuristic algorithm based on Harris Hawks Optimization (HHOLS) for Task Scheduling in Fog Computing (TSFC) in Industrial Internet of Things (IIoT) applications. The algorithm is designed to improve the Quality of Services (QoSs) provided to users through cloud computing. It integrates a swap mutation operation and a local search strategy to improve its performance and considers energy consumption, makespan, cost, flow time, and carbon dioxide emission rate (CDER) as performance metrics.

A task scheduling algorithm for fog computing platforms, with the aim of reducing the total system makespan and energy consumption presented in [38]. The proposed approach consists of two key components: 1) an Ant Mating Optimization (AMO) bio-inspired optimization approach and 2) optimized distribution of tasks among the fog nodes within proximity. Performance evaluation results in this paper demonstrates that the proposed approach provides better makespan and energy consumption than the bee life algorithm (BLA), traditional PSO and GA. Furthermore [40], a multi-objective evolutionary algorithm, namely M-E-AWA, has been proposed for task scheduling in cloud computing environments. This algorithm focuses on optimizing transmission time, processing time, and resource utilization with the aim of reducing response time and energy consumption costs. Experimental results demonstrate that M-E-AWA effectively utilizes complex Pareto fronts and outperforms comparative methods, such as Round Robin (RR) and simple genetic algorithms, in terms of execution time and service cost metrics.

TABLE II
COMPARISON BETWEEN REVIEWED PAPERS WITH OUR PROPOSED METHOD, CT: COMPLETION TIME, MS:MAKESPAN, EC: ENERGY CONSUMPTION, WT: WAIT TIME, D: DEADLINE, FT: FAULT TOLERANCE, S: SIMULATOR

| | Game theory | CT/MS | EC | WT | D | FT | S |
|---|---|---|---|---|---|---|---|
| [4] | - | - | + | - | - | - | iFogsim |
| [17] | + | + | - | - | - | - | ✕ |
| [22] | - | + | - | - | - | - | ✕ |
| [23] | - | + | + | - | + | - | iFogsim |
| [28] | - | - | + | - | - | - | ✕ |
| [30] | - | + | + | - | - | - | Java |
| [38] | - | - | + | + | - | - | Matlab |
| [40] | - | + | - | - | - | + | Matlab |
| GAP | + | + | + | + | + | + | iFogsim |

## III. SYSTEM MODEL

Our objective system design is based on the fog computing environment and contains $N=\{1,2,...,n\}$ set of users, $M=\{1,2,...,m\}$ set of fog nodes, $T=\{1,2,...,t\}$ set of tasks, $R$ for reliability, and $P$ for power consumption. We employ non-cooperative games to strike a balance between reliability and power usage within the designated time frame in fog computing. Our aim is to attain exceptional reliability while minimizing power consumption, all while adhering to the specified deadline. A visual representation of the system model in the form of a flowchart can be observed in Fig. 3. Initially, we will present the recommended approach, followed by an in-depth discussion on power and reliability management in Section 4. The suggested strategy effectively harmonizes power consumption and reliability by dynamically adjusting the system's operational frequency and voltage. This method empowers the system to operate at a reduced power level while still meeting the required deadline and maintaining high reliability. To illustrate, in the context of an innovative urban application, fog nodes can be strategically positioned throughout various locations to provide services such as traffic control, environmental monitoring, and rescue operations. Multiple fog nodes can be assigned to each task to ensure maximum reliability. Users can access these services using their mobile devices, and the system can adaptively allocate resources based on the availability of fog nodes and task priorities. To minimize power consumption, we enhance the work allocation and scheduling algorithms.

### A. Power Model

In digital systems, the total power consumption consists of two components: static power $P_{static}$ and dynamic power $P_{dynamic}$. The dynamic power consumption depends on the activity of the system [13][25]. To calculate the dynamic power consumption, we utilized Equation (1) [13][34].

$$P_{dynamic}(V_i, f_i) = \alpha . C_L V_i^2 f_i \qquad (1)$$

where $\alpha$ is equal to the activity, and $C_L$ is equal to the total capacity of the internal load capacitors of the circuit. Also, $V_i^2$ is equal to the supply voltage and $f_i$ is equal to the working frequency. In our proposed method, the DVFS technique is used to reduce dynamic power consumption, which this technique does by shifting the voltage level and frequency [13][33].

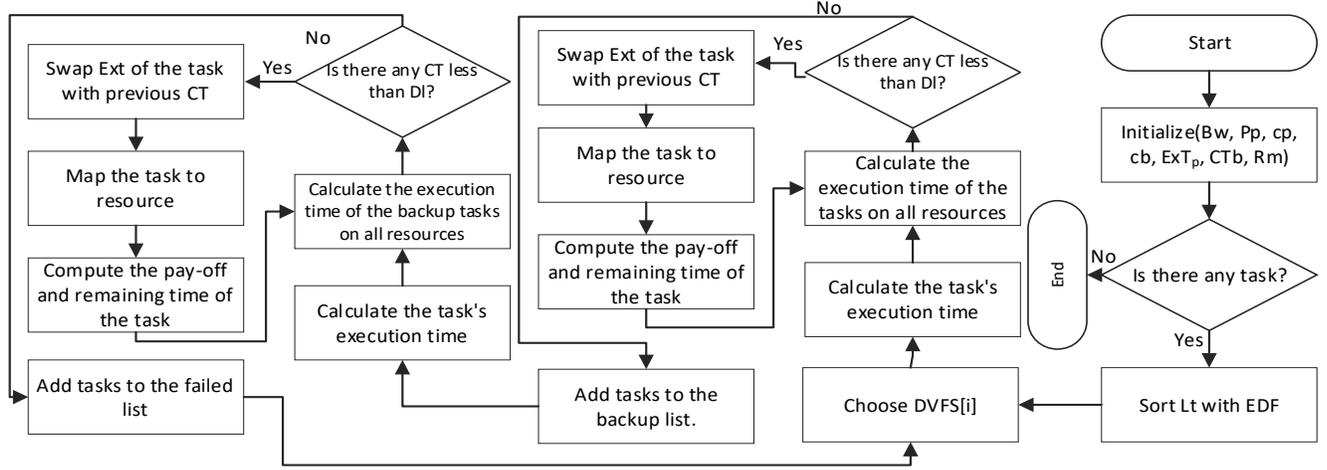

**Fig. 3.** Visual representation of the different steps of GAP algorithm for scheduling tasks

### B. Reliability Model

Considering a system's power and energy consumption, evaluating its efficiency requires careful consideration of its reliability. The role of the reliability model in our proposed methodology is to ensure that the system not only operates efficiently in terms of power and energy usage but also maintains consistent and accurate performance over time. Reliability modeling helps in identifying and mitigating potential faults that could disrupt service, thereby enhancing the overall robustness of the system. One practical approach to enhance the system's reliability involves employing a primary/backup (PB) system configuration. This system's primary and backup components can be implemented in various ways. For instance, in this scenario, the primary and backup tasks may reside on the same or different nodes, with the backup task operating in a lower power state. In the event of a primary component failure, the backup component assumes control, and if the backup also encounters a failure, the entire system becomes non-operational [24].

We have incorporated the CPB technique into the system to model fault tolerance. This technique defines the primary unit as the fundamental service unit, while the backup unit is a secondary entity that can be selected to take over when necessary. When the primary unit experiences a fault, the backup unit initiates its operation, and the execution time for each task can be determined using equation (2):

$$Exec_{T_i} = PT_i(t) + BT_i(t) \qquad (2)$$

where $PT$ represents the primary task and $BT$ the backup task, the backup unit ensures system functionality even in the event of primary task failure. System reliability is then calculated using equation (3), as detailed in [21] and [39]:

$$R(t) = e^{-\lambda t} \qquad (3)$$

where, the variable $\lambda$ represents the occurrence rate of faults, while $t$ denotes the time. Equation (4), as refered in [21], demonstrates the occurrence rate of defects based on frequency.

$$\lambda(f) = \lambda_0 . 10^{\frac{d(1-f)}{1-f_{min}}} \qquad (4)$$

In this equation, $\lambda_0$ is equal to the average fault rate at the maximum frequency and $d$ is equal to the sensitivity factor. The sensitivity factor is a measure of how the transient faults change when voltage and frequency are scaled.

## IV. AN OVERVIEW OF GAP METHODOLOGY

The previous methods focused solely on task scheduling and resource allocation without considering reliability. In contrast, our proposed model, which incorporates non-cooperative games, addresses fault tolerance in scheduling through the CPB technique. High power consumption leads to a decrease in reliability. Fig. 4 illustrates the basic structure of the model.

The CPB technique is vital in managing reliability and fault tolerance in our proposed method. This approach schedules two identical tasks on separate nodes to ensure non-overlapping execution. Consequently, in error-free scenarios where tasks must be completed without errors, the system maintains low overhead in terms of time [14][20].

Utilizing game theory, our proposed model encompasses scheduling, reliability management, and power consumption control. It involves end-user requests, a scheduler, and the available resources of fog computing nodes. Algorithm 1 outlines how the GAP model balances reliability and power considerations in a cloud-fog environment.

### A. Problem Definition

In this section, we define the problem of task scheduling in fog computing environments, which refers to static scheduling problems. Tasks have been scheduled non-preemptively, meaning they will execute until completion.

The goal of a scheduling algorithm should be to maximize the amount of work finished in a given amount of time rather than to select an execution plan that minimizes execution time.

To model the scheduler, non-cooperative game theory has been used. In a scheduling problem that may arise in any real system of two or more competing agents, the agents (or players) may want to execute their tasks simultaneously. However, from the point of view of the system as a whole, when more than one task needs to be completed and any delay in completion delays all the tasks, the deadline for the task may be missed, causing low reliability. It is likely to benefit the system by concentrating resources on these tasks. In the proposed method, users (end devices) are the players, and the strategies that users can employ include sending or not sending the task to resources. The players' strategy is represented: $S = [s_i(t_0), s_i(t_1), ..., s_i(t_n)]$. To achieve the objective as mentioned above, equation (5) has been used.

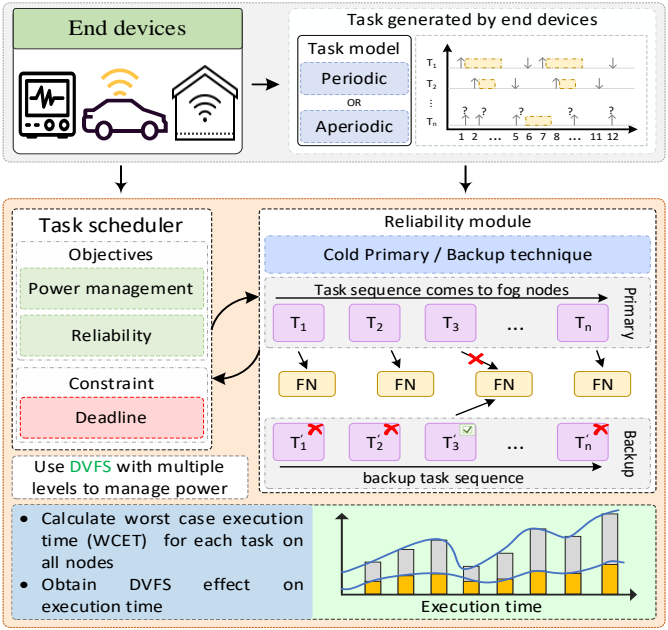

**Fig. 4.** This diagram depicts a task scheduler using DVFS for power management and a CPB technique for reliability, balancing execution time and energy efficiency for periodic and aperiodic tasks.

$$\forall i \text{ if } x_{ij} = 1: Min \sum_{j=1}^{m} P_j \left( \sum_{i=1}^{t} t_{ij} . Exec_i \right) \quad (5)$$

This ensures that tasks are mapped to resources in a way that minimizes total power consumption while considering execution times. We want to make a trade-off between power consumption and reliability using non-cooperative game theory by designing a scheduler that is a bi-objective problem, which is challenging to solve because they involve two or more conflicting objectives. We first analyzed the characteristics of the problem and observed that it is a bi-objective problem, which can be solved using the GAP method. This problem has two main parts: node assignments and the required voltage-frequency level of each of the primary and backup tasks needed to meet the desired result. To correctly assign tasks to nodes and achieve the desired result, we must find the voltage-frequency level of all tasks.

A bi-objective problem is a single problem whose optimal solution requires optimizing two or more arbitrary variables [26]. Finally, we formulated the task scheduling problem using the GAP method.

Because of the GAP algorithm's task deadline constraint, tasks with shorter deadlines must take precedence and are prioritized over tasks with more extended deadlines. Tasks with high priority need resources that must execute before the deadline. Low-priority tasks may need different resources than high-priority ones. To map the resources and tasks, we use the following mathematical expression:

$$\forall t_i: Min(d_i) \rightarrow Max(r_j) \quad (6)$$

We assign the task with the nearest deadline to the resource with the maximum available capacity. Therefore, we choose the solution that will provide desirable results. In this case, Fig. 5 shows some of the resources that may not be chosen to execute tasks by our game-theoretic algorithm, and some resources may have more than one task at the same time. These tasks are carefully evaluated and chosen by the game-theoretic algorithm to ensure that a beneficial solution is obtained.

**Objectives:** The GAP algorithm's core objective is to schedule the tasks to minimize power consumption while maintaining deadline compliance and reliability target. The algorithm chooses a method for scheduling the tasks' execution so that no deadline is missed. To minimize power consumption, the DVFS technique has been used. Total power consumption is shown in equation (7). The sum of the power used by all the tasks makes up the total power used by the system.

$$P_{Total} = \sum_{i=1}^{T} P_i(V_{max}, f_{max}) \quad (7)$$

Equation (10) represents the power consumption after applying the DVFS technique to the algorithm to minimize the power consumption. When the DVFS technique has been applied the voltage and frequency will scale.

$$V_i = \rho_j V_{max}, \qquad 0 < \rho_j < 1 \quad (8)$$

---

**Algorithm 1** GAP Algorithm

    **Input**: $Bw$, $Pp$, $Rt$
    **Output**: List (Task), List (Resources)
    **Initialization**: assign VMs {$Bw$, $Pp$}; $cp$, $cb = 0$, 0;
    $CTb[] =$ **null**, $ExT_P = 0$, $Rm[] =$ **null**
1.  **if** ($LoT$.size() != **null**) **then**
2.     $Lt :=$ sort($LoT$, $LoT_{dl}$)
3.     **for** *DVFS* **do**
4.         **for** all tasks $t_i$ in $Lt$ **do**
5.             $LCT := ExT_{i0}$
6.         **for** all resources $r_j$ in $LR$ **do**
7.             **if** $CT_i > ExT_{ij}$ **do**
8.                 $ExT_P := Tr$;
9.                 $LCT_i := ExT_{ij}$;
10.                 **map** $t_i$ to $r_{ij}$; compute pay-off ($t_i, r_{ij}$);
11.                 $Rm_i := ExT_{ij} - ExT_P$;
12.         **Repeat** until all tasks are mapped
13.         **if** $t_i$ not executed **do** $LT_b[] := t_i$;
14.             $c_p$++;
15.     $Lt_b :=$ sort( $LT_b$, $Rm$[index of $t_i$])
16.     **for** all tasks $t_i$ in $Lt_b$ **do**
17.         $LCT_i := ExT_{i0}$
18.     **for** all resources $r_j$ in $LR$ **do**
19.         **if** $Rm[t_i] > ExT_{ij}$ **and not in same** vm **do**
20.             **map** $t_i$ to $r_{ij}$;
21.             compute pay-off ($t_i, r_{ij}$);
22.         **Repeat** until all tasks in $Lt_b$ are mapped
23.         **if** $t_i$ not executed **do** $L_f[] := t_i$;
24.             $c_b$++;
25.         **end**
26.     **end**
27.     **end**
28.  **end**

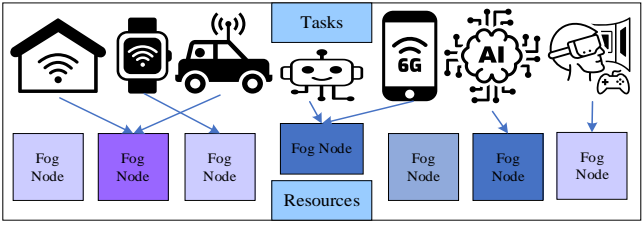

**Fig. 5.** An example of how tasks can be organized and assigned to the available resources.

$$f_i = \rho_i f_{max}, \qquad 0 < \rho_i < 1 \qquad (9)$$

$$Minimize\ P_{Total} = \sum_{i=1}^{T} P_i\ (V_i, f_i) \qquad (10)$$

The DVFS technique could cause low reliability. The voltage and frequency should have specific levels and be set properly. Because the fault rate λ depends on voltage level, by decreasing voltage, the fault rate λ increase exponentially, according to equation (3). As a result, the voltage Vi fault rate will be presented [27][34]:

$$\lambda(V_i) = \lambda_0 . 10^{\frac{V_{max} - V_i}{d}} \qquad (11)$$

Where $\lambda_0$ is the average fault rate at the maximum voltage and $d$ is equal to the sensitivity of the component to scaled voltage. The voltage side, as expressed in equation (11), and the frequency side, as expressed in equation (4), are both critical components in assessing the reliability of a system. Their influence on the system's stability should not be underestimated, as even small changes can have drastic repercussions. By carefully examining both equations and their effects, it is possible to ensure the reliability and longevity of any system.

**Constraints:** In the proposed method, the primary restriction is the deadline of tasks, which is a time constraint. The completion time of each task must not exceed the given deadline. Equation (12) introduces a binary variable, $x_{ij}$, which indicates whether task $i$ is assigned to resource $j$. The variable is 1 if the task is assigned, and 0 otherwise.

$$\forall_i\ if\ x_{ij} = 1: c_i \le d_i \qquad (12)$$

### B. Algorithm Discussion

The proposed approach for solving the scheduling problem revolves around prioritizing tasks with lower power consumption while ensuring a nearly stable reliability level and successful completion of all tasks. The algorithm consists of two phases: the first phase focuses on executing the main task, while the second phase handles backup tasks. In the background, a power management technique and an algorithm for computing pay-offs are employed to select the optimal model.

Algorithm 1 presents the pseudo-code for the scheduling algorithm. The initial lines (1-3) declare input, output, and parameter initialization for the pseudo-code. In lines 1-2, task presence is checked, and the tasks are sorted based on their estimated deadlines using the "Earliest Deadline First" (EDF) approach in this section. Line 3 adjusts the frequency utilized by the fog devices. A comparison is then made among different frequency levels to determine the best frequency that enables the fog devices to consume less power while maintaining satisfactory reliability.

Lines 4-9 calculate the execution times of all tasks on all available resources, selecting the resource with the shortest execution time for each task. Lines 10-12 involve mapping the task selected in the previous section to a resource and computing the task's pay-off or reward by running the task with another resource before its deadline. If the task still has remaining time for execution, it is saved; otherwise, it is automatically set to 0. This process is repeated until all tasks have been assigned to resources. If a task cannot be mapped to a resource or encounters a fault during execution, it is added to the backup task list to be handled in lines 13-14. Line 15 organizes the tasks not completed in the first phase, sorting them based on the time remaining and the deadline for resource completion.

Lines 18-22, similar to lines 4-10, calculate the execution time for the backup tasks. However, the resources used for the backup tasks possess greater computing power than those used in the first phase.

Ultimately, if a task fails to execute, it is considered a failure and added to the failure list in line 23. The number of failed tasks is counted in line 24. The DVFS technique is employed to adjust frequency levels and strike a balance between objectives, precisely the number of failed tasks, or the need for running backup tasks. In other words, tasks that experience faults are compared, and their reliability is computed.

### C. Algorithm Complexity Analysis

The GAP algorithm guarantees to converge to O(nlog(n)) time complexity. The worst-case execution time (WCET) of the proposed method is O(nlog(n)). The WCET of the algorithm occurs where all content of the DVFS list is tracked on all primary and backup tasks. The heap sort algorithm is used to sort the primary and backup tasks. The time complexity of this algorithm in the worst-case scenario is O(nlog(n)), and its space complexity is O(n) [36]. Heap sort is the only sorting algorithm that has both O(nlog(n)) time complexity with O(1) for space complexity together because no extra data structure has been used in it [36]. In the proposed algorithm, mapping the task to the resources with a specific frequency takes the most time.

## V. RESULTS AND DISCUSSION

In this section, the GAP methodology is experimented with and evaluated in the iFogsim simulation tool because there is no control in a real-world environment. iFogsim [35] is a tool that experiments can repeat multiple times with different configurations. For evaluating our GAP methodology, we employ the experimental setup shown in Fig. 6.

### A. Experimental Setup

In this study, we showcase the efficiency of the GAP method using iFogsim by evaluating its impact on power consumption, energy consumption, and execution time. The specific configuration employed in iFogsim for conducting the simulation using the GAP methodology is outlined in Table II.

To assess the effectiveness of our approach, we compare it against seven other algorithms, measuring their performance using three distinct metrics. The proposed algorithm generates a list of tasks (LoT) with random lengths ranging from 1000 to 2000, and a list of resources (LoR) with random MIPS values

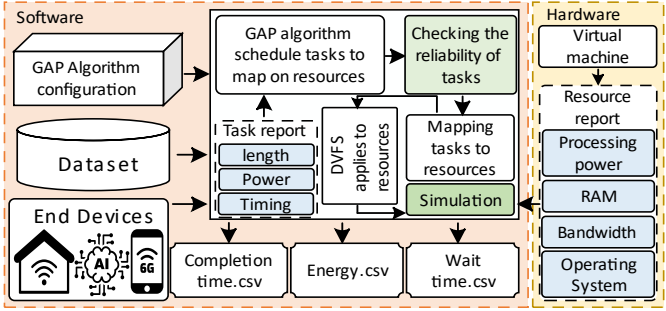

**Fig. 6.** The experimental setup

### TABLE IV
### TASKS AND RESOURCES CONFIGURATION

| Parameter | Value |
|---|---|
| Task Length | [1000, 2000] |
| Task Amount | [200, 400, 600, 800, 1000] |
| Task NPE | [1, 8] |
| VM MIPS | [1000, 2000] |
| VM RAM | 256 Mb |
| VM BW | 1000 B/s |
| VM Amount | [20, 50, 80, 100] |
| VM NPE | [1, 8] |

between 1000 to 2000. The performance of our method positions it as a promising candidate for practical implementation. The outcomes indicate that the proposed approach outperforms existing methods, providing a more efficient solution. The evaluation criteria are defined as follows: 1) energy consumption, 2) reliability, 3) completion time, 4) wait time, and 5) power consumption.

Table III shows tasks and resources configuration like task length, Number of Processor Elements (NPE), resource RAM, number of tasks and VMs, etc.

### B. Energy Consumption

Overall, the energy consumption of a system is the total amount of energy used by its components, such as fog nodes and sensors. This can be measured by calculating the power consumed over a specific period using the formula ($P \times t$), where $P$ represents the power consumed, and $t$ represents the time.

This method provides an easy way to determine the energy consumed within a specific time frame. It can help identify areas where energy consumption can be optimized or reduced. It is also essential to consider the efficiency of the system's components and the overall system itself. This can be done by comparing the energy consumed to the system's work. Below, in Fig. 7, we consider different utilizations and compare the energy consumption of our proposed method with seven other algorithms.

Fig. 7a illustrates that the proposed method efficiently consumes energy. With its maximum number of tasks, GAP consumes less energy than the other seven algorithms in any number of tasks (key observation ①). We did not use DVFS in WGAP, and the results of WGAP are better than the four other algorithms. After using the DVFS technique, GAP consumed

### TABLE III
### HOST CONFIGURATION PARAMETERS

| Parameter | Value |
|---|---|
| Architecture | X86 |
| Bandwidth | 10000 B/s |
| RAM | 2048 Mb |
| Storage | 100000 Mb |
| OS | CentOS |
| VM model | Xen |
| Time Zone | 8.0 |
| Cost | 2 |
| Cost per memory | 0.01 |
| Cost per storage | 0.001 |

54%, 56%, 67%, and 80% less than the First Come First Serve (FCFS), Shortest Job First (SJF), Round Robin (RR), and Particle Swarm Optimization (PSO) algorithms. The PSO algorithm consumed significantly more energy due to the local minimum or the number of iterations required to identify the most effective task for node mapping.

One observation in Fig. 7b demonstrates that GAP uses energy effectively in each case. Interestingly, even without DVFS, GAP could outperform the four other algorithms regarding energy efficiency (key observation ②). This implies that GAP could handle its operations more efficiently than the other four algorithms. It was well-suited to the task and could operate with fewer resources while producing the same results. Furthermore, the performance of GAP was not affected by the local minimums or the number of iterations required to find the optimal task for node mapping, as it maintained its energy efficiency. This shows that GAP has a remarkable advantage over the other algorithms regarding energy efficiency.

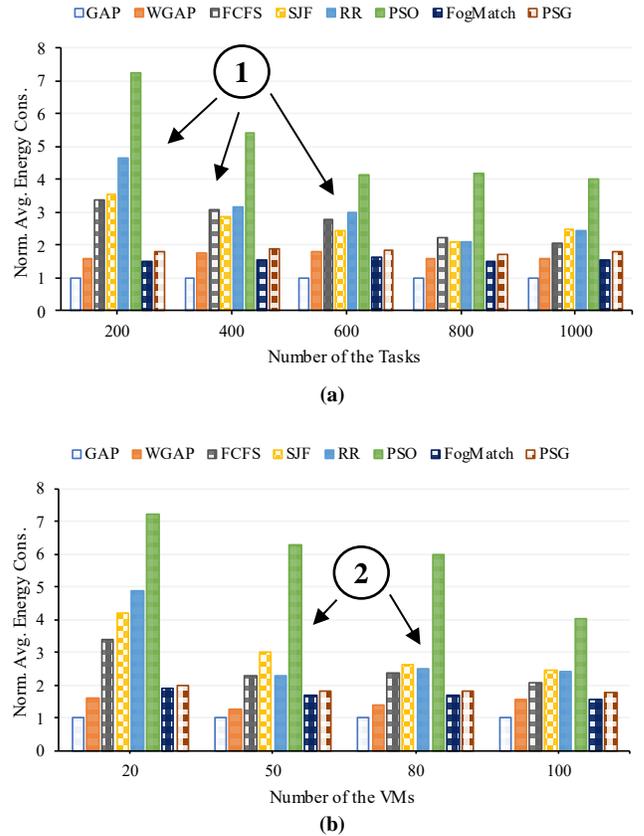

**Fig. 7.** The tasks' normalized Energy consumption (lower is better for energy efficiency). The results show the efficiency of the GAP method in combination with DVFS. a) The comparison of the 7 methods showed energy consumption for handling 100 resources, b) The comparison of the 7 methods showed energy consumption for handling 1000 tasks.

Additionally, the results of the experiment showed that GAP was able to maintain its high level of energy efficiency even when the workload increased. This is particularly important in real-world scenarios where the workload can vary significantly over time. In contrast, the other algorithms showed a decrease in energy efficiency as the workload increased, which could lead to higher energy costs and longer processing times. Overall, the findings of this study demonstrate the effectiveness of the GAP algorithm in reducing energy consumption and improving performance in distributed computing environments

### C. Reliability Analysis

In our proposed method, not only the power and energy consumption have been considered, but also the reliability of the method itself is considered an important metric. A reliable system is more efficient and cost-effective in the long run and crucial for ensuring the safety and success of any application it uses. In our approach, we have made a trade-off that all of the tasks execute without any failed tasks, and on the other hand, we have used the DVFS technique with low power and energy consumption. Also, we execute all of the tasks.

Fig. 7 shows that the energy used by the other and the WGAP

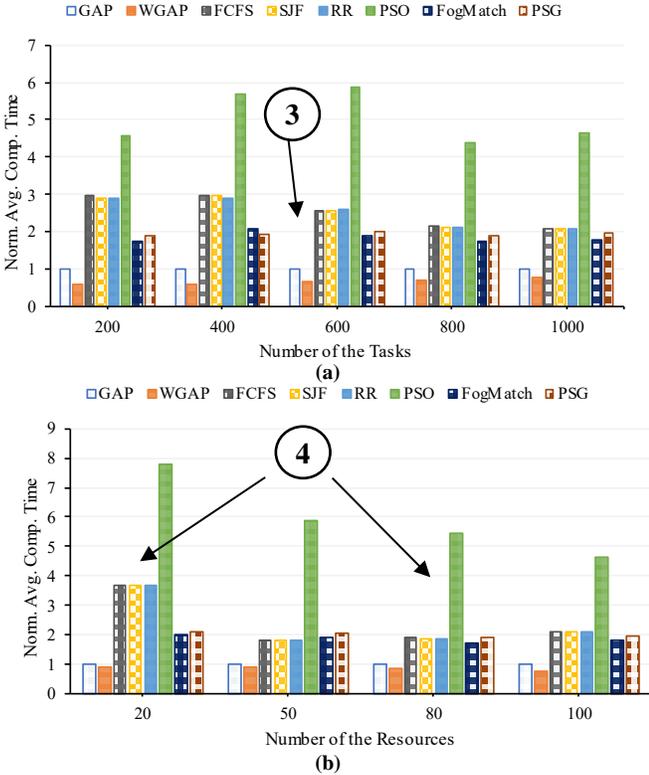

**Fig. 8.** The normalized completion time of tasks on fog nodes (lower is better). a) Seven different methods were tested to see which could complete 100 resources in the shortest amount of time, b) A comparison of the 1000 tasks' completion times for each of the seven methods revealed that method WGAP was the most efficient.

methods without DVFS is higher than the GAP algorithm. However, the overhead of the GAP method is reflected in the wait and completion times. Despite this, the GAP method's wait time and completion time are still better than those of the FCFS, SJF, RR, and PSO algorithms by at least 41%.

### D. Completion Time

Completion time is a standard metric in usability testing [37]. It measures the time a user takes to perform a task (from start to finish). Fig. 8a and 8b show the total completion time of the tasks with different utilizations.

The following mathematical expression depicts the average completion time of the tasks:

$$ACT = \frac{\sum_{i=0}^{n} CT_{ij}}{N} \tag{13}$$

$$CT = ST_i + Exec_{ij} \tag{14}$$

The results shown in Fig. 8 demonstrate the superior performance of the GAP algorithm in terms of average completion time (ACT). The proposed method's ACT is superior to the other algorithms in every scenario, but the WGAP has less completion time (key observation ③). In the worst-case scenario, the ACT of the GAP algorithm is even better than the best-case scenario for the FCFS or the other three algorithms. This indicates that the GAP algorithm can promptly and effectively complete tasks, making it a reliable and efficient option for various applications. Overall, the results indicate that the GAP algorithm is highly effective in timely completing tasks. Its average completion time is consistently superior to other algorithms, even in the worst-case scenario. This makes it a reliable and efficient option for a variety of applications.

Figs. 8a and 8b demonstrate that the WGAP algorithm has a shorter completion time than GAP (key observations ③). However, when the DVFS technique is employed, the GAP algorithm's completion time is extended, in comparison to WGAP (key observations ④). Ultimately, the GAP algorithm is the quickest of all four algorithms.

### E. Wait Time

The wait time for tasks with deadlines is an important metric because the task may miss its deadline if it increases. The wait time is a crucial metric for evaluating a task scheduling algorithm. The GAP method outperforms the others in minimizing wait time for task deadlines. This is particularly important in time-sensitive industries such as manufacturing or healthcare. By reducing the wait time, the scheduling algorithm can ensure that tasks are completed on time, minimizing delays and improving overall efficiency. As Fig. 9b shows, the wait time decreases as more fog nodes are added (key observations ⑥) due to the high availability of resources. This is also reflected in the completion rate, proving that more fog nodes lead to faster task execution (key observations ④). Fig. 9 shows the proposed method's average wait time (AWT) compared to seven others. The wait time of seven different methods for 1000 tasks can vary drastically depending on the number of resources available (see Fig. 9a). We devised equation (16) to calculate the average wait time for the tasks.

$$W_t = s_{ij} - ST_i \tag{15}$$

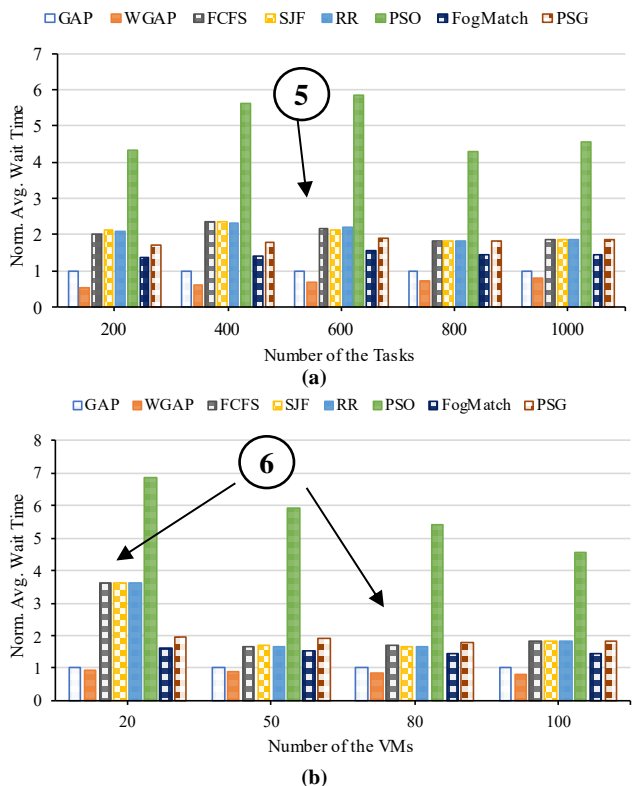

**Fig. 9.** The normalized wait time of tasks on fog nodes (lower is better). a) Contrasting the wait times of seven distinct approaches for 100 resources, b) An analysis of the wait times for 1000 tasks using each of the seven methods using various numbers of resources revealed significant differences in efficiency. WGAP consistently had the shortest wait times across all resource levels, while PSO had the longest wait times.

$$AW_t = \frac{\sum_{i=0}^{n} W_t}{n} \qquad (16)$$

To show the impact of the wait time of the algorithm, we considered multiple utilities; these represent the fact that a low wait time causes a low completion time of the GAP compared to others. Compared to other algorithms that are implemented, the GAP algorithm has at least 31% low wait time, making that algorithm more efficient.

Furthermore, the AWT of the GAP algorithm remains consistently low even as the number of tasks and deadlines increase. This indicates that the algorithm can efficiently schedule tasks and prioritize those with closer deadlines. However, it is essential to note that AWT is not the only metric to consider when evaluating task scheduling algorithms. Other factors, such as task completion time, should also be considered. As seen in Fig. 8a and Fig. 9a, the WGAP method has a lower wait time and completion time than GAP (key observations ③,⑤) because of the nature of the DVFS algorithm used in the GAP.

### G. Average Power Consumption

Average Power Consumption (AWC) is the average amount of power used by a device or system over a certain period of time. This value is usually measured in watts (W) or kilowatts (kW), as shown in Figures 10a and 10b, where the vertical axis represents power consumption in kilowatt-hours (kWh).

In simpler terms, if we know how much power a device or system uses in an hour, that is its average power consumption.

As seen in Figures 10, the GAP algorithm initially uses more power than other algorithms. However, the GAP, FogMatch, and PSG algorithms show relatively stable power usage across different task loads, ranging from 5 to 7 units. In general, the WGAP and PSO algorithms perform worse than the GAP algorithm. The WGAP algorithm consistently shows an increase in power usage as the number of tasks increases, starting at around 3 units and reaching around 7 units. The main reason for the high power usage of the WGAP algorithm is that it uses maximum voltage and frequency to execute all tasks. It is worth mentioning that the FogMatch and PSG algorithms, as discussed in previous studies [36] and [88], show lower average power consumption than the GAP algorithm, but their longer execution times result in higher energy consumption and fewer successfully executed tasks. The FCFS, SJF, RR, and PSO algorithms show minor fluctuations but generally remain around the 6-unit mark.

Figure 10b illustrates the AWC of various scheduling algorithms as the number of resources increases. It is clear that SJF consistently shows the lowest power consumption, making it the most energy-efficient algorithm. In contrast, WGAP shows the highest power consumption, indicating potential scalability issues. GAP, FCFS, RR, PSO, FogMatch, and PSG exhibit stable or slightly decreasing power consumption trends, with FogMatch and PSG performing better at higher VM counts. These insights highlight SJF's superiority in energy efficiency and the need to optimize

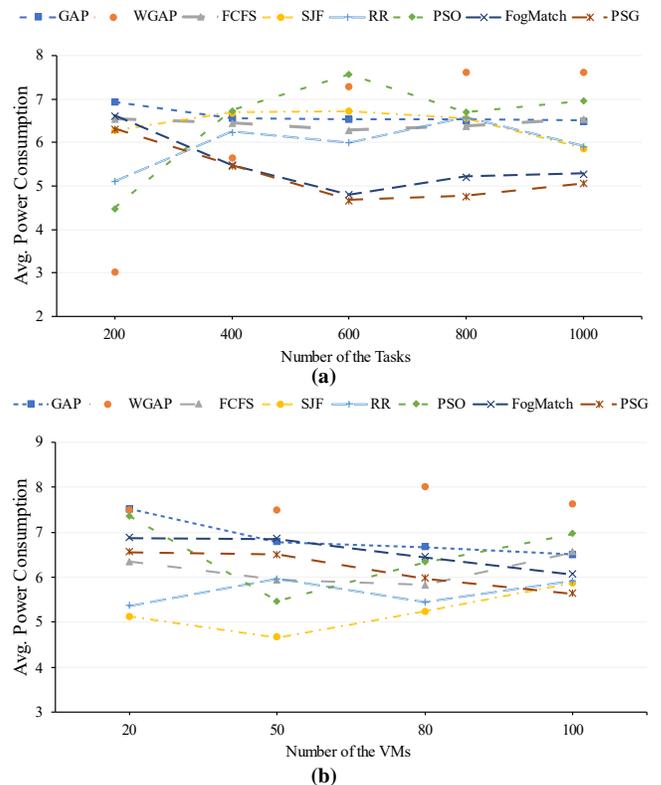

**Fig. 10.** highlights the energy efficiency of scheduling algorithms, with SJF consuming the least power and WGAP the most. GAP, FogMatch, and PSG algorithms maintain stable power usage across different task loads, showcasing their scalability.

WGAP for better scalability in energy-sensitive environments, which becomes GAP.

## V. CONCLUSION

This study explores a method that utilizes non-cooperative game theory to balance power consumption and reliability in task scheduling problems. The method presented in this paper aims to determine an optimal task-resource allocation strategy that minimizes power usage and the occurrence of task failures, all while maintaining high overall performance. Notably, our algorithm prioritizes tasks based on their deadlines and assigns resources based on the required processing power for each task. The approach adopted by the algorithm demonstrates how the suggested approach outperforms the standard baseline algorithms. Though this paper aims to develop a more efficient scheduling approach, it needs to make more use of traditional theoretical algorithms. The simulation results affirm the effectiveness of the GAP algorithm in achieving both low power consumption and stable, reliable task execution through efficient trade-offs in task scheduling. Additionally, our method considers the heterogeneity of resources and tasks, making it applicable to various computing systems. The algorithm also considers the dynamic nature of task scheduling problems, adapting to changes in the system's workload and resource availability.

In future work, we plan to further explore the integration of machine learning techniques to enhance the scheduling algorithm's efficiency and accuracy. Additionally, we aim to conduct experiments on real-world computing systems to evaluate the practicality and effectiveness of the proposed approach. Another avenue for future research is to investigate the impact of task dependencies on the scheduling algorithm and develop strategies to optimize task execution order. Overall, the proposed GAP algorithm has the potential to revolutionize task scheduling in computing systems, improving performance and reducing energy consumption.

## ACKNOWLEDGEMENT

This work was supported in part by the NYUAD's Center for CyberSecurity (CCS), funded by Tamkeen under the NYUAD Research Institute under Grant G1104.

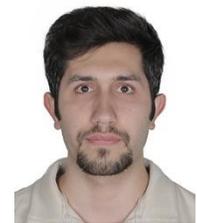

**Abolfazl Younesi** received his M.Sc. degree in computer engineering from Sharif University of Technology (SUT), Tehran, Iran, in 2024. He is currently a member of the Embedded Systems Research Laboratory (ESR-LAB) at the department of computer engineering, Sharif University of Technology. His research interests include the Internet of Things (IoT) and Cyber-Physical Systems (CPS), low-power design, machine learning, and computer vision.

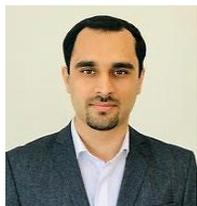

**Mohsen Ansari** is currently an assistant professor of computer engineering at Sharif University of Technology, Tehran, Iran. He received his Ph.D. degree in computer engineering from Sharif University of Technology, Tehran, Iran, in 2021. He was a visiting researcher in the Chair for Embedded Systems (CES), Karlsruhe Institute of Technology (KIT), Germany, from 2019 to 2021. He is currently the director of Cyber-Physical Systems Laboratory (CPSLab) at Sharif University of Technology. He was the technical program committee (TPC) member of ASP-DAC (2022 and 2023). Dr. Ansari is serving as an associate editor of the IEEE Embedded Systems Letter (ESL). His research interests include embedded machine learning, low-power design, real-time systems, cyber-physical systems, and hybrid systems design.

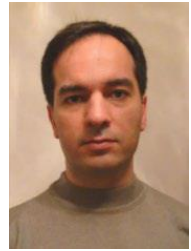

**Alireza Ejlali** received the Ph.D. degree in computer engineering from the Sharif University of Technology (SUT), Tehran, Iran, in 2006, where he is currently an Associate Professor of computer engineering. From 2005 to 2006, he was a Visiting Researcher with the Electronic Systems Design Group, University of Southampton, Southampton, U.K. He is currently the Director of the Embedded Systems Research Laboratory with the Department of Computer Engineering, Sharif University of Technology. His research interests include low power design, real-time, and fault-tolerant embedded systems.

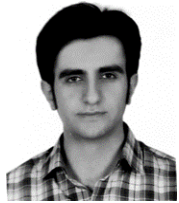

**MohammadAmin Fazli** received his B.Sc. in hardware engineering and MSc and PhD in software engineering from Sharif University of Technology, in 2009, 2011 and 2015 respectively. He is currently a Lecturer at Sharif University of Technology and R&D Supervisor at Sharif's Intelligent Information Center resided in this university. His research interests include Game Theory, Combinatorial Optimization, Computational Business and Economics, Graphs and Combinatorics, Complex networks and Dynamical Systems.

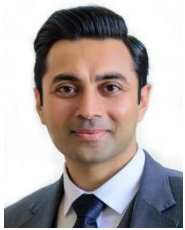

**Muhammad Shafique** (Senior Member, IEEE) received the Ph.D. degree in computer science from the Karlsruhe Institute of Technology (KIT), Germany, in 2011. Afterwards, he established and led a highly recognized research group at KIT for several years and conducted impactful collaborative research and development activities across the globe. In October 2016, he joined the Faculty of Informatics, Institute of Computer Engineering, Technische Universität Wien (TU Wien), Vienna, Austria, as a Full Professor in computer architecture and robust, energy-efficient technologies. Since September 2020, he has been with New York University (NYU) Abu Dhabi, United Arab Emirates, where he is currently a Full Professor and the Director of the eBrain Laboratory and a Global Network Professor at the Tandon School of Engineering, NYU-New York City, USA. He is also a Co-PI/an Investigator in multiple NYUAD Centers, including the Center of Artificial Intelligence and Robotics (CAIR), the Center of Cyber Security (CCS), the Center for InTeractIng urban nEtworkS (CITIES), and the Center for Quantum and Topological Systems (CQTS). His research interests include AI & machine learning hardware and system-level design, brain-inspired computing, quantum machine learning, cognitive autonomous systems, wearable healthcare, energy-efficient systems, robust computing, hardware security, emerging technologies, FPGAs, MPSoCs, and embedded systems. His research has a special focus on cross-layer analysis, modeling, design, and optimization of computing and memory systems. The researched technologies and tools are deployed in application use cases from Internet of Things (IoT), smart cyber-physical systems (CPS), and ICT for development (ICT4D) domains. He has given several keynotes, invited talks, tutorials, and organized many special sessions at premier venues. He has served as the PC chair, the general chair, the track chair, and a PC member for several prestigious IEEE/ACM conferences. He holds one U.S. patent, has coauthored six books, more than ten book chapters, more than 350 papers in premier journals and conferences, and more than 50 archive articles. He received the 2015 ACM/SIGDA Outstanding New Faculty Award, the AI 2000 Chip Technology Most Influential Scholar Award, in 2020 and 2022, the ASPIRE AARE Research Excellence Award, in 2021, six gold medals, and several best paper awards and nominations at prestigious conferences. He is a Senior Member of IEEE Signal Processing Society (SPS) and a member of the ACM, SIGARCH, SIGDA, SIGBED, and HIPEAC.

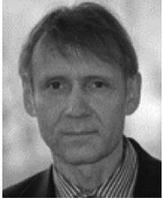

**Jörg Henkel** (Fellow, IEEE) received the diploma and PhD (summa cum laude) degrees from the Technical University of Braunschweig, Germany. He was a research staff member with NEC Laboratories, Princeton, NJ. His research work is focused on co-design for embedded hardware/soft-ware systems with respect to power, thermal, and reliability aspects. He has received six best paper awards from, among others, ICCAD, ESWeek, and DATE. He served as the editor-in-chief for the ACM Transactions on Embedded Computing Systems and IEEE Design&Test. He is/has been an associate editor for major ACM and IEEE journals. He was a general chair ICCAD, ESWeek, etc., and serves as a Steering Committee chair/member for leading conferences and journals. He coordinates the DFG Program SPP 1500 "Dependable Embedded Systems" and is a site coordinator of the DFG-TR89 collaborative research center on "Invasive Computing." He is the Chairman of the IEEE Computer Society, Germany Chapter.

APPENDIX. A

**Lemma 1 (Optimality of Task-Resource Mapping):** Let $T$ be the set of tasks, $R$ be the set of resources and $M$ be the set of possible mappings between tasks and resources. Let $P$ be the pay-off function that maps each task-resource pair to a real value. Let $LCT_i$ be the latency of task $i$, and $ExT_{ij}$ be the expected execution time of task $i$ on resource $j$. Let $Rm_i$ be the remaining resources available for task $i$. Then, the algorithm's mapping of tasks to resources is optimal if and only if it satisfies the following conditions:

$$\forall i \in T, \forall j \in R, \quad LCT_i = \min_{k \in R} E \, xT_{ik} \Rightarrow P(i,j) \ge P(i,k)$$

$$\forall i \in T, \quad \sum_{j \in R} P(i,j) \le Rm_i \tag{1}$$

**Lemma 2 (Correctness of Latency Computation):** Let $LCT_i$ be the latency of task $i$, and $ExT_{i0}$ be the expected execution time of task $i$ on the initial resource. Then, the algorithm correctly computes the latency of each task if and only if:

$$LCT_i = ExT_{i0} \Rightarrow LCT_i = \min_{k \in R} E \, x T_{ik} \tag{2}$$

**Lemma 3 (Correctness of Resource Allocation):** Let $Rm_i$ be the resources available for task $i$, and $ExT_{ij}$ be the expected execution time of task $i$ on resource $j$. Then, the algorithm correctly allocates resources to tasks if and only if:

$$Rm_i = ExT_{ij} - ExT_p \Rightarrow \sum_{j \in R} P(i,j) \le Rm_i \tag{3}$$

*Theorem 1: Correctness of the Algorithm*

Let $LoT$ be the list of tasks, $LoT_{dl}$ be the deadline list, $LR$ be the list of resources, and $Bw$, $Pp$ be the bandwidth and processing power of the virtual machines, respectively. Let $CTb$ be the array of completion times, $ExT_p$ be the expected total processing time, and $Rm$ be the array of remaining resources. Then, the algorithm correctly maps tasks to resources, computes the pay-off for each task-resource pair, and allocates resources to tasks.

**Proof:** We first show that the algorithm correctly sorts the tasks based on their deadlines. Let $Lt$ be the sorted list of tasks. Then, by construction, we have:

$$Lt = \text{sort}(LoT, LoT_{dl}) \Rightarrow \forall i, j \in Lt, \, i < j \Rightarrow LoT_{dl_i} \le LoT_{dl_j} \tag{4}$$

We then show that the algorithm correctly computes the latency of each task. Let $LCT_i$ be the latency of task $i$. Then, by Lemma 2, we have:

$$LCT_i = ExT_{i0} \Rightarrow LCT_i = \min_{k \in R} E \, xT_{ik} \tag{5}$$

We then show that the algorithm correctly maps tasks to resources. Let $t_i$ be a task, and $r_j$ be a resource. Then, by construction, we have:

$$CT_i > ExT_{ij} \Rightarrow ExT_p = T_t; \quad LCT_i = ExT_{ij};$$
$$\text{map } t_i \text{ to } r_j; \quad \textbf{compute pay-off}\big(t_i, r_{ij}\big) \tag{6}$$

By Lemma 1, this mapping is optimal.

We then show that the algorithm correctly allocates resources to tasks. Let $Rm_i$ be the remaining resources available for task $i$. Then, by Lemma 3, we have:

$$Rm_i = ExT_{ij} - ExT_p \Rightarrow \sum_{j \in R} P(i,j) \le Rm_i \tag{7}$$

Finally, we show that the algorithm correctly computes the pay-off for each task-resource pair. Let $P(i, j)$ be the pay-off for task $i$ and resource $j$. Then, by construction, we have:

$$P(i,j) = \textbf{compute pay-off}\big(t_i, r_{ij}\big) \Rightarrow P(i,j) \ge P(i,k) \, \forall k \in R \tag{8}$$

**Corollary 1 (Optimality of the Algorithm):** The algorithm is optimal in the sense that it maximizes the total pay-off of the tasks.

**Proof:** By Lemma 1, the algorithm's mapping of tasks to resources is optimal. By construction, the algorithm computes the pay-off for each task-resource pair and allocates resources to tasks based on the pay-off. Therefore, the algorithm maximizes the total pay-off of the tasks.

*Theorem 2: Convergence of the Algorithm*

**Proof:** By construction, the algorithm iteratively maps tasks to resources and allocates resources to tasks based on the payoff. By Lemma 1, the algorithm's mapping of tasks to resources is optimal. Therefore, the algorithm converges to an optimal solution in a finite number of iterations.

**Corollary 2 (Robustness of the Algorithm):** The algorithm is robust to changes in the input parameters, such as the deadlines and processing powers of the tasks and resources.

**Proof:** By construction, the algorithm is based on a sorting step, which is robust to changes in the input parameters. The algorithm's mapping of tasks to resources and allocation of resources to tasks are based on the pay-off, which is a function of the input parameters. Therefore, the algorithm is robust to changes in the input parameters.

*Theorem 3: Scalability of the Algorithm*

The algorithm is scalable to large inputs, such as a large number of tasks and resources.

**Proof:** By construction, the algorithm has a time complexity of O(n log n) and a space complexity of O(n), where n is the number of tasks. Therefore, the algorithm is scalable to large inputs.

**Corollary 3 (Flexibility of the Algorithm):** The algorithm is flexible and can be easily modified to accommodate different scheduling policies and constraints.

**Proof:** By construction, the algorithm is based on a modular design, where each step is independent of the others. Therefore, the algorithm can be easily modified to accommodate different scheduling policies and constraints. In conclusion, we have shown that the algorithm is correct, optimal, convergent, robust, scalable, and flexible. The algorithm's design is based on a rational and complex mathematical framework, which ensures that it is theoretically sound and practically applicable.